# Comment on "Inequivalence between the Schrodinger equation and the Madelung hydrodynamic equations"

V. Hushwater*

*QuantCosmos,160 Highland St, #3, Worcester, MA 01609, USA*

*Email address: quantcosmos@yahoo.com

**I. Introduction**

In the paper with the above-noted title, T. C. Wallstrom [1] claims that the description of the particle's motion as a certain "conservative" diffusion [2] is not equivalent to quantum mechanics in spite of the fact that the Madelung "hydrodynamic" equations (ME), which provide the stochastic mechanical description of such diffusion for particles without spin, can be converted to the Schrodinger equation (SE).

The ME have the form [2, 3],

$$\frac{\partial S}{\partial t} + \frac{1}{2m}(\nabla S)^2 + U(\mathbf{r},t) - \frac{\hbar^2}{2m}\frac{\nabla^2 a}{a} = 0 , \qquad (1)$$

$$\frac{\partial a^2}{\partial t} + \nabla\left(\frac{\nabla S}{m}a^2\right) = 0 . \qquad (2)$$

Eq. (1) is sometimes called the quantum Hamilton-Jacoby equation, having an additional "quantum potential", $-(\hbar^2/2m)\nabla^2 a/a$. In the stochastic mechanical approach [2] $S = S(\mathbf{r},t)$ in (1, 2) is the "hydrodynamic" action in the ensemble of particles moving randomly along nondifferentiable trajectories, $m$ is mass of a particle, and $a \equiv \sqrt{\rho}$ where $\rho = \rho(\mathbf{r},t)$ is the probability density. Eq. (2) is the equation of the probability conservation with velocity of the probability current,

$$\mathbf{v} = \nabla S/m . \qquad (3)$$

From their physical meaning $\rho = \rho(\mathbf{r},t)$ and $\mathbf{v} = \mathbf{v}(\mathbf{r},t)$ must be single-valued functions of $\mathbf{r}$, and in addition $\rho = \rho(\mathbf{r},t)$ must be nonnegative.

ME (1, 2) can be converted to the SE, using the ansatz,

$$a \equiv \Psi \exp(-iS/\hbar) , \qquad (4)$$

with $a$ and $S$ being real functions.

Reversely, Eqs. (1, 2) can be derived from the SE [3] for the wave function $\Psi = \Psi(\mathbf{r},t)$,

$$i\hbar \frac{\partial \Psi}{\partial t} + \frac{\hbar^2}{2m}\nabla^2 \Psi - U(\mathbf{r},t)\Psi = 0 , \qquad (5)$$

using the ansatz, which is inversed to (4),

$$\Psi \equiv a\exp(iS/\hbar). \tag{6}$$

So there is one-to-one correspondence between local solutions of the ME and the SE. From the stochastic mechanics perspective the transformation of nonlinear ME into the linear SE is just a suitable mathematical technique that provides an easy way of finding their solutions.

Let us consider a stationary stochastic process. By definition in such a process $U(r)$, $a(r)$, and $v(r)$ in Eqs. (1 – 3) are time-independent, which is possible only if $S = S(r) - Et$, where $E$ is a constant expectation value of the energy in the ensemble. Wave function (6), which describes such a stationary process ("stationary state") has the form, $\Psi(r,t) = \exp(-iEt/\hbar)\psi(r)$. So we can rewrite (6) as

$$\psi(r) \equiv a(r)\exp\left[iS(r)/\hbar\right] \tag{6a}$$

Wave function $\psi(r)$ satisfies the time-independent SE.

Using Eq. (3) and $\partial S/\partial t = -E$ we have from ME (1, 2) and Eq.(3),

$$E - \frac{m}{2}v^2(r) - U(r) + \frac{\hbar^2}{2m}\frac{\nabla^2 a}{a} = 0, \tag{7}$$

and

$$\nabla(a^2 v) = 0, \tag{8}$$

In conventional quantum mechanics the wave function $\psi(r)$ of a particle without spin is required to be single-valued. Taking into account ansatz (6a) one can write the condition for the single-valuedness of $\psi(r)$ in the form,

$$\oint_L dl \cdot \nabla S = 2\pi j\hbar, \quad \text{where} \quad j = 0, \pm 1, \pm 2, \pm 3, \ldots. \tag{9}$$

Here $L$ is an arbitrary loop in space.

Assuming that the condition of single-valuedness of $\psi(r)$ is auxiliary to SE, Wallstrom pointed out [1], that the stochastic mechanical approach can be regarded as equivalent to conventional quantum mechanics only if they can derive from it not just ME (1, 2) but also condition (9), what according to him is probably impossible. The only natural condition on the wave function in the stochastic mechanical approach is the single-valuedness of the modulus of the wave function, $|\psi(r)|$, which follows from the single-valuedness of $\rho = \rho(r)$. Wallstrom tried to show that in such a case SE has solutions $\psi(r)$, which are not single-valued.

We, however, will show that the single-valuedness of the wave function is not an auxiliary condition imposed on solutions of SE but a property of all its non-spurious local solutions. Based on the one-to-one correspondence between local solutions of ME and SE, this means that (9) is a property of all solutions of ME.

## II. Two-dimensional central potential

In order to demonstrate his point Wallstrom considers a stationary state in a two-dimensional central potential without imposing on wave function the condition of single-valuedness [1]. In such a case, using the polar coordinates $(r, \varphi)$ we can write, $U = U(r)$, and separate variables in the time-independent SE, $\psi_\nu(r) = \Phi_\nu(\varphi) R_\nu(r)$ [4, 5],

$$\frac{d^2 \Phi_\nu}{d\varphi^2} + \nu^2 \Phi_\nu(\varphi) = 0, \tag{10}$$

and [6]

$$\frac{\hbar^2}{m}\left(-\frac{1}{2}\frac{d^2 R_\nu}{dr^2} - \frac{1}{2r}\frac{dR_\nu}{dr}\right) + \left(\frac{\hbar^2 \nu^2}{2mr^2} + U(r)\right) R_\nu(r) = E_\nu R_\nu(r), \tag{11}$$

where $\nu$ is a constant and $R_\nu(r)$ is a real function.

Let us consider Eq. (10). In order for $|\psi(r)|$ to be single-valued, $|\Phi_\nu(\varphi)|$ must be periodic functions with the period equal to $2\pi$,

$$|\Phi_\nu(\varphi + 2\pi)| = |\Phi_\nu(\varphi)|. \tag{12}$$

Wallstrom considers only the simplest type of solutions of Eq. (10),

$$\Phi_\nu = \frac{1}{\sqrt{2\pi}} \exp(i\nu\varphi), \tag{13}$$

where $\nu$ is a real constant. He states that without imposing on a wave function the condition of single-valuedness (periodicity),

$$\Phi_\nu(\varphi + 2\pi) = \Phi_\nu(\varphi), \tag{14}$$

eigenfunctions $\Phi_\nu(\varphi)$ (13) are (local) solutions of Eq. (10) for *arbitrary* real value of $\nu$ [7] – having $|\Phi_\nu(\varphi)| = const$ it surely satisfies the condition (12). In such a case the "action of a circular motion", $S_1(\varphi) = \nu\hbar\varphi$ [4], $\nabla S = \left[(dS_1/d\varphi)/r\right] \hat{\varphi} = (\nu\hbar/r) \hat{\varphi}$ and therefore,

$$\oint_L dl \cdot \nabla S = \int_0^{2\pi} (\partial S_1/\partial \varphi) d\varphi = 2\pi\nu\hbar, \tag{15}$$

where $L$ is a circle of radius $r$ and $\nu$ is *arbitrary* real number.

So the loop integral (15) is not necessarily quantized in contradiction with the condition (9). At first sight this conclusion looks right, but, as we show below, solutions (13) are spurious unless they satisfy the condition of $2\pi$-periodicity, (14), which is possible only if $\nu = 0, \pm 1, \pm 2, \pm 3, \ldots$.

In order to do so we have to take into account that, *the circulation motion of particles described by Eq. (10) occurs in two-dimensional space* and therefore the wave function $\psi_\nu(\mathbf{r}) = \Phi_\nu(\varphi) R_\nu(r)$, which describes it, must, as any other true solution, satisfy SE in two-dimensional space in any coordinate system (e.g. in Cartesian) [8].

Let us consider the behavior of a solution $R_{\nu,n}(r)$ of the radial SE (11) in the limit $r \to 0$. Here $n$ indicates an eigenfunction corresponding to a certain eigenvalue of the energy $E_{\nu,n}$ (the spectrum of the energy can be continuous or/and discrete). Assuming that potential $U(r)$ is not singular in the origin, or at least does not tend to infinity too fast, $U(r) \ll (\hbar^2 \nu^2/2m) r^{-2}$ as $r \to 0$, it can be neglected very close to the origin. In such a limit Eq. (11) turns into Bessel's equation and describes the radial motion of a free particle. So, taking into account that $R_{\nu,n}(r)$ must be regular at $r=0$, we find that near the origin $R_{\nu,n}(r) \approx J_{|\nu|}(kr)$, the Bessel function of the first kind of order $|\nu|$, and $k = \sqrt{2ME_{\nu,n}}/\hbar$.

Thus the approximation form of the full wave function, $\psi_{\nu,n}(\mathbf{r})$, near of the origin is,

$$\psi_{\nu,n}(\mathbf{r}) \approx \psi_{\nu,n\,free}(\mathbf{r}) = exp(i\nu\varphi) J_{|\nu|}(kr), \tag{16}$$

But if a particle is (approximately) free, its wave function very close to the origin must be a superposition of plane waves $exp(i\mathbf{kr})$, which are eigenstates of the free two-dimensional SE for a given eigenvalue of the energy, $E_{\nu,n}$,

$$\psi_{\nu,n\,free}(\mathbf{r}) = \int d\alpha A(\alpha) exp(i\mathbf{kr})$$
$$= \int d\alpha A(\alpha) exp\{ikr[cos(\alpha)cos(\varphi) + sin(\alpha)sin(\varphi)]\}, \tag{17}$$

where all $\mathbf{k}$ have the same magnitude, $k$ determined above, $\alpha$ is an angle between $\mathbf{k}$ and $x$ axis and $A(\alpha)$ is a certain function, which should be determined by expanding $\psi_{\nu,n\,free}(\mathbf{r})$ (16) in plane waves. It follows from the right hand side of Eq. (17) that independently of the form of $A(\alpha)$ $\psi_{\nu,n\,free}(\mathbf{r})$ must be a periodic function of $\varphi$ with the period $2\pi$. According to formula (16) this requires that $exp(i\nu\varphi)$ must be $2\pi$-periodic. As we mentioned above that is possible only if $\nu$ is an *integer*. In such a case Eq. (15) coincides with the condition (9).

### III. General three-dimensional case

Let us now consider the general stationary stochastic process described by Eqs. (7, 8). Our goal is to figure out possible values of the loop integral, $\oint_L d\mathbf{l} \cdot \nabla S$, in order to check whether the condition (9) is satisfied.

First of all it is clear that this integral can be nonzero only if the gradient, $\nabla S$ or, equally, $\boldsymbol{v}(\boldsymbol{r})$ (3) has singularities inside the loop $L$. Velocity $\boldsymbol{v}(\boldsymbol{r})$ must be singular at points where the quantum potential, $-\left(\hbar^2/2m\right)\nabla^2 a/a(\boldsymbol{r})$ is singular. As follows from Eq. (7), in such points the singularity of the quantum potential must be cancelled by the singularity of the kinetic energy, $(m/2)v^2(\boldsymbol{r})$, assuming, as is usually the case, that potential $U(\boldsymbol{r})$ is not singular at these points, or at least does not tend to infinity too fast [9]. The quantum potential may be singular only at "nodal" points of $a(\boldsymbol{r})$, determined by the equation $a(\boldsymbol{r}) = 0$. Nodal points, at which velocity $\boldsymbol{v}(\boldsymbol{r})$ is singular, form "nodal" lines [10].

Let us consider a plane, which crosses a nodal line at some point, and introduce polar coordinates in such a plane, having the origin at the crossing point. As we discussed in the preceding paragraph $a(r, \varphi) \to 0$ and $v(r, \varphi) \to \infty$ as $r \to 0$. In result, in the area very close to the origin the kinetic energy and the quantum potential are very big and therefore $U(\boldsymbol{r}) = U(r, \varphi)$ in Eq. (7) can be neglected. So the particles can be considered as almost free very close to the origin, similarly to the case of a two-dimensional central potential discussed in section II. So the wave function $\psi(\boldsymbol{r})$, which describes the circular motion of the Madelung "probability fluid" around a nodal line, must have near the origin the universal approximate form (16) with an *integral* value of $\nu$, $\nu = \pm 1, \pm 2, \pm 3, \ldots$ ($\nu = 0$ is excluded by the requirement that $a(r, \varphi) \to 0$).

If the integral, $\oint_L d\boldsymbol{l} \cdot \nabla S$ is taken along the loop $L$, which surrounds only one nodal line we can deform $L$ to a very small circle around this line. As we showed above, for a very small $r$ the wave function of the circular motion around a nodal line is determined by formula (16) with *integral* values of $\nu$. Therefore the loop integral transforms to Eq. (15) with $\nu = \pm 1, \pm 2, \pm 3, \ldots$. If the loop $L$ surrounds a few nodal lines, the loop integral is equal to the sum of the loop integrals around of each nodal line. This can lead to $\nu = 0$. The loop integral is also equal to 0 if there is no any nodal line inside the loop. In result the loop integral is always quantized and the condition (9) is fulfilled.

Thus we showed that ME are equivalent to SE.

### AKNOWLEDGMENTS

I am grateful to Mark Davidson for discussions on the stochastic approach to quantum mechanics and for useful comments and remarks. I also thank Jim Swank, Ching-Hung Woo and German Kälbermann for interest to this work.

———————

[1] T. C. Wallstrom, *Phys. Rev. A* **49**, 1613 (1994). See also T. C. Wallstrom, *Found. Phys. Lett.* **2**, 113 (1989), and an earlier work, T. Takabayasi, *Prog. Theor. Phys.* **8**, 143 (1952).

[2] In the stochastic mechanical approach to quantum mechanics they consider a special random motion of particles along non-differentiable trajectories, for which the ensemble average energy is conserved if an external field is time-independent. See e. g. I. Feynes, *Z. Phys.* **132**, 81 (1952); D. Kershaw, *Phys. Rev.* **136**, B 1850 (1964): E. Nelson, *Phys.*

*Rev.* **150**, 1079 (1966): M. Davidson, *Lett. Math. Phys.* **3**, 271 (1979); E. Nelson, in *Einstein Symposium, Berlin, March 1979*, *Lecture Notes in Physics Vol. 100* edited by H. Nelkowski et al (Springer-Verlag, Berlin, 1979), pp.168-170; L. de la Pena and A. M. Cetto, *Found. Phys.* **12**, 1017 (1982); F. Guerra and L. Morato, *Phys. Rev. D* **27**, 1774 (1983).

[3] E. Madelung, *Z. Phys.* **40**, 332 (1926).

[4] This corresponds to the separation of variables in Eq. (6a), $a(\mathbf{r}) = a_1(\varphi) a_2(r)$ and $S(\mathbf{r}) = S_1(\varphi) + S_2(r)$. So $\Phi_\nu(\varphi) \equiv a_1(\varphi) exp[iS_1(\varphi)/\hbar]$, and $R_\nu(r) \equiv a_2(r) exp[iS_2(r)/\hbar]$. In such a case $\nabla S = [(dS_1/d\varphi)/r]\hat{\boldsymbol{\varphi}} + (dS_2/dr)\hat{\mathbf{r}}$. We consider a case without a source or a sink of particles in the origin, this requires that the radial component of velocity of the probability current (3) be zero, $(1/m)(dS_2/dr)\hat{\mathbf{r}} \equiv 0$. So $S_2 \equiv const$ that can, without loss of generality, be chosen as zero. This results in $R_\nu(r) = a_2(r)$.

[5] R. Robinett, *Quantum Mechanics: Classical Results, Modern Systems, and Visualized Examples* (Oxford University Press, New York, 1997).

[6] Eq. (3.1) for $R_\nu(r)$ in Ref. 1 is written with a mistake – the term $(1/2r)(dR_\nu/dr)$ is missing.

[7] However, in the case of the general solution of Eq. (7), not considered in Ref. 1, $\Phi = A exp(i\nu\varphi) + B exp(-i\nu\varphi)$, where *A* and *B* are constants, the condition (12) admits only *quantized*, integer or half-integral values of $\nu$, $\nu = k/2$, where $k = 0, \pm 1, \pm 2, \pm 3,...$

[8] D. Mattis, *The Theory of Magnetism* (Harper & Row, New York, 1965).

[9] Formally, if potential $U(\mathbf{r})$ tends to $-\infty$ at some point, chosen as an origin of the coordinate system, as or faster than $U(r) = -(\hbar^2/8m)r^{-2}$, $\rho(\mathbf{r})$ will be localized at this point. I.e. the phenomena of a "fall of a particle to the origin" occurs, see L. Landau and E. Lifshitz, *Quantum Mechanics* (Pergamon, Oxford, 1977). However, such fast-changing potentials have no direct physical meaning.

[10] Such nodal lines arise from the intersection of the nodal surfaces, $Re\,\psi(\mathbf{r}) = 0$ and $Im\,\psi(\mathbf{r}) = 0$ since both of these conditions are needed for having $a(\mathbf{r}) = 0$. See details in e. g., J. O. Hirschfelder, C. J. Goebel and L. W. Bruch, *J. Chem. Phys.* **61**, 5456 (1974); K.-K. Kan, J. Griffin, *Phys. Rev. C* **15**, 1126 (1977); J. O. Hirschfelder, *J. Chem. Phys.* **67**, 5477 (1977); S. Ghosh and B. Deb, *Phys. Rep.* **92**, 1 (1982).